\documentclass{article}
\usepackage{spconf,amsmath,amssymb,graphicx,dsfont}
\usepackage{amsfonts}

\usepackage{soul,color}
\usepackage[latin1]{inputenc}
\usepackage{pgfplots, subfigure}

\DeclareMathOperator*{\argmin}{argmin}
\DeclareMathOperator*{\argmax}{argmax}

\newcommand{\ie}{\emph{i.e., }}

\newcommand{\Rr}{\mathds{R}}
\newcommand{\Cr}{\mathds{C}}

\newcommand{\D}{\mathbf{D}}

\newcommand{\xv}{\mathbf{x}}

\newcommand{\yv}{\mathbf{y}}

\newcommand{\dv}{\mathbf{d}}

\newcommand{\rv}{\mathbf{r}}

\newcommand{\tv}{\text{\boldmath$\theta$}}

\newcommand{\Ns}{\mathcal{N}}

\newcommand{\Os}{\mathcal{O}}

\title{Phase recovery from a Bayesian point of view:\\ the variational approach}
\name{Ang\'elique Dr\'emeau, Florent Krzakala}
\address{\normalsize{Sorbonne Universités, UPMC Univ Paris 06, UMR
    8550, LPS-ENS F-75005, Paris,    France}\\
\normalsize{Laboratoire de Physique Statistique de l'\'Ecole Normale Sup\'erieure, 29
rue Lhomond, 75005, Paris,    France}}

\begin{document}
%
\maketitle
\begin{abstract}
  In this paper, we consider the phase recovery problem, where a complex signal
  vector has to be estimated from the knowledge of the modulus of its
  linear projections, from a naive variational Bayesian point of
  view. In particular, we derive an iterative algorithm following the minimization of
  the Kullback-Leibler divergence under the mean-field assumption, and
  show on synthetic data with random projections that this approach
  leads to an efficient and robust procedure, with a reasonable computational
  cost.
 \end{abstract}

\section{Introduction}
\label{sec:intro}

Reconstructing a complex vector given only the magnitude of
measurements, a problem known under the name of \emph{phase recovery
  problem}, is at the heart of numerous applications, such as
crystallography \cite{Harrison1993}, coherent diffractive
\cite{Quiney2010} or optical \cite{Katz2014} imaging. Formally, the
problem writes as follows: observing $\yv\in\Rr_+^M$ through a
$M\times N$ complex-measurement matrix $\D$, we aim at recovering the
signal $\xv\in\Cr^N$, such that
\begin{align}
\yv=|\D\xv|.\label{eq:basic}
\end{align}
It is a non-convex optimization problem notoriously difficult to solve. Therefore, many algorithms have been devised in the literature to deal with this problem. 
We can roughly divide them into two main families.
\begin{enumerate}
\item \emph{The alternating-projection algorithms} alternate projections on the span of the measurement matrix and on the object domain. Among these approaches, we can mention the works of Gerchberg \& Saxton \cite{Gerchberg1972}, Fienup \cite{Fienup1982} and Griffin \& Lim \cite{Griffin1984}. 
\item \emph{The algorithms based on convex relaxations} propose to approximate the phase recovery problem by relaxed problems which can be solved efficiently by standard optimization procedures. Two of the main approaches of this type, namely \emph{PhaseLift} \cite{Chai2011, Candes2013} and \emph{PhaseCut} \cite{Waldspurger2013}, rely in particular on semidefinite programming. 
\end{enumerate}
In an attempt to circumvent the non-linearity of the modulus, a
Bayesian point of view was recently adopted in \cite{Schniter2012}.
Focusing on the particular context of compressive measurements, the
approach exploits sparse priors and resorts to a
"loopy-belief-propagation'' type algorithm
\cite{murphy1999loopy,mezard2009information} in a continuation of a
line of works in compressive sensing
\cite{Donoho2010,Rangan2011,Krzakala2012}.


Here instead, we release the compressive constraint and consider the
case where the vector of interest is a full, non-sparse vector.  We
follow the naive variational Bayesian approach based on a mean-field
approximation. We show that, as long as our simulation setup is
concerned, the proposed algorithm is competitive with state-of-the-art
procedures, while emphasizing some desirable properties, namely an
higher robustness to noise and a reasonable computational complexity.

\section{Bayesian point of view}
\label{sec:statement}
In this section, we recall the Bayesian modeling introduced in
\cite{Schniter2012}, which we shall follow in the remaining of the
paper, and introduce the estimation problem we propose to solve.

\subsection{Model}
\label{subsec:model}
Reinterpreting problem \eqref{eq:basic} into a Bayesian framework, we introduce new variables, modeling, on the one hand, the missing phases of the observations, and on the other hand, some acquisition noise. Thus, each absolute-valued measurement
$y_\mu$, $\mu\in\lbrace 1\ldots M\rbrace$ of $\yv$ is expressed as
\begin{align}
y_\mu =e^{j\theta_\mu} \big(\sum_{i=1}^N x_{i}\;d_{\mu i} +n_\mu\big),\label{eq:y}
\end{align}
where $\theta_\mu\in[0,2\pi)$ stands for its missing conjugate phase,
and $n_\mu$ is a zero-mean circular Gaussian noise with variance $\sigma_n^2$. 
We suppose moreover that
\begin{align}
&p(\xv) = \prod_{i=1}^N p(x_i) \quad\quad \text{with} \quad\quad p(x_i)=\mathcal{C}\Ns(0,\sigma_x^2),\label{eq:x}\\
\text{and}&\nonumber\\
&p(\tv) = \prod_{\mu=1}^M p(\theta_\mu) \quad\quad \text{with} \quad\quad  p(\theta_\mu) = \frac{1}{2\pi}.\label{eq:th}
\end{align}
Under these assumptions, the absence of phases in the observations is naturally taken into account in the model since marginalizing on $\theta_\mu$ leads to a distribution on $y_\mu$ which only depends on the moduli of $y_\mu$ and $\sum_{i=1}^N x_{i}\;d_{\mu i}$.

\subsection{Problem formulation}
\label{subsec:pb}
Within model \eqref{eq:y}-\eqref{eq:th}, the recovery of the complex signal of interest $\xv$ can be expressed as the solution of the following marginalized Maximum A Posteriori (MAP) estimation problem
\begin{align}
\hat{\xv} &= \argmax_\xv p(\xv|\yv),\label{eq:pb}\\
\text{with}\quad\quad &p(\xv|\yv) = \int_\tv p(\xv,\tv|\yv).
\end{align}
Because of the marginalization on the hidden variables $\tv$, the direct computation of $p(\xv|\yv)$ is intractable. We can however envisage different sub-optimal techniques to solve \eqref{eq:pb}. In this paper, we focus on the solutions brought by variational approaches and derive the procedure related to a particular one: the mean-field approximation.

\section{Variational approximations}
\label{sec:approx}

\subsection{General formulation}
\label{subsec:KL}
Keeping in mind our previous notations, the variational approximations aim at approximating the posterior joint distribution $p(\xv,\tv|\yv)$ by the distribution $\hat{q}(\xv,\tv)$ leading to the minimum of the Kullback-Leibler (KL) divergence conditionally to a set of given constraints $\mathcal{F}$:
\begin{align}
\hat{q}(\xv,\tv) &= \argmin_{q\in\mathcal{F}} \int_\xv\int_\tv q(\xv,\tv) \log \big(\frac{q(\xv,\tv)}{p(\xv,\tv|\yv)} \big) d\xv \; d\tv. \label{eq:KL}
\end{align}
Depending on $\mathcal{F}$, the minimization \eqref{eq:KL} gives raise to different approximations \cite{Yedidia2005}. We mention here two of them.
\begin{itemize}
\item With $\mathcal{F} = \big\lbrace q\big| q=\prod_{i=1}^N q_i(x_i)\prod_{\mu=1}^M q_\mu(\theta_\mu)\big\rbrace$, we define a Mean-Field (MF) approximation and problem \eqref{eq:KL} can be efficiently solved using the ``Variational Bayes Expectation-Maximization" (VB-EM) algorithm \cite{Beal2003}. 
\item With $\mathcal{F} = \big\lbrace q\big| q=\frac{\prod_{a=1}^A q_a(\xv_a)\prod_{b=1}^B q_b(\tv_b)}{\prod_{i=1}^N q_i(x_i)^{\alpha_i-1}\prod_{\mu=1}^M q_\mu(\theta_\mu)^{\beta_\mu-1}}\big\rbrace$ where $[\xv_1\ldots \xv_A]$ (resp. $[\tv_1\ldots \tv_B]$) partitions the variables $\xv$ (resp. $\tv$) and $\alpha_i$ (resp. $\beta_\mu$) is the degree of variable node $x_i$ (resp. $\theta_\mu$), problem \eqref{eq:KL} refers to the minimization of the Bethe free energy, which can be solved by 
generalized approximate message passing (GAMP) algorithms \cite{Rangan2011,6875083}. This is the approach followed in \cite{Schniter2012}.
\end{itemize}
To the best of our knowledge, the MF approximation has never been
considered in a phase retrieval context, within model
\eqref{eq:y}-\eqref{eq:th}. We detail the structure of the resulting
algorithm in the next subsection. With respect to the more involved
Bethe approach, the variational MF one has some clear
advantages: it is conceptually much simpler and leads immediately to a
fast iterative algorithm, without further approximation. It is also
more mathematically grounded as it leads to a {\it provably convergent
  algorithm}, and to a guaranteed {\it provable bound} of the KL
divergence (this is not the case of the Bethe approach
\cite{wainwright2008graphical}), all that, we shall see, with very
similar performances (at least in the non-sparse cases we are
considering here).
 
\subsection{Variational Bayes EM algorithm}
\label{subsec:MF}
The VB-EM algorithm is an iterative procedure which successively updates the factors of the considered MF approximation. Particularized to model  \eqref{eq:y}-\eqref{eq:th} and the MF approximation defined above, it leads to the following update equations\footnote{For a sake of clarity, we drop here the iteration indices.}:
\begin{align}
q(\theta_\mu) &=  \frac{1}{2\pi I_0(\frac{2}{\sigma_n^2}|y_\mu^*\langle z_\mu\rangle|)}\;\exp\left(\frac{2}{\sigma_n^2} \Re(y_\mu^*\langle z_\mu\rangle e^{j\theta_\mu})\right),\label{eq:qt}\\
q(x_i) &= \mathcal{C}\Ns(m_i,\Sigma_i),\label{eq:qx}
\end{align}
where
\begin{align}
&\Sigma_i = \frac{\sigma^2_n\sigma^2_x}{\sigma^2_n+ \sigma_x^2\dv_i^H\dv_i}, \\
&m_i = \frac{\sigma_x^2}{\sigma_n^2+ \sigma_x^2\dv_i^H\dv_i} \langle\rv_i\rangle^H\dv_i, \label{eq:moy2}\\
&\langle \rv_i \rangle=\bar{\yv}-\sum_{k\neq i} m_k\;\dv_k,\label{eq:resi2}
\end{align}
with
\begin{align}
&\bar{\yv}=\left [y_\mu e^{(j \arg(y_\mu^*\langle z_\mu\rangle))}\;\frac{I_1(\frac{2}{\sigma_n^2}|y_\mu^*\langle z_\mu\rangle|)}{I_0(\frac{2}{\sigma_n^2}|y_\mu^*\langle z_\mu\rangle|)}\right]_{\mu=\lbrace1\ldots N\rbrace}\\
& \langle z_\mu\rangle = \sum_{i=1}^N m_i\; d_{\mu i}.
\end{align}
In the equations above, $I_0$ (resp. $I_1$) stands for the modified Bessel function of the first kind for order $0$ (resp. $1$), $.^H$ denotes the conjugate transpose and $.^*$ the scalar conjugate.

At this point of discussion, it can be interesting to compare the proposed procedure and the sparse-decomposition algorithm proposed in \cite{Dremeau2012}. Although solving different problems (phase recovery for the one, sparse decomposition for the other), the algorithms rely both on MF approximations, and share indeed notable structural similarities.
\begin{itemize}
\item[\emph{i)}]  In particular, \eqref{eq:qx} - \eqref{eq:resi2} are identical in both algorithms. The only difference lies in the definition of the vector $\langle \rv_i \rangle$. More than a simple current residual, $\langle \rv_i \rangle$ takes here into account complex-valued observations $\bar{\yv}$, in which the ``reconstructed" phase $ \arg(y_\mu^*\langle z_\mu\rangle)$, reminding of the MAP estimate of $\theta_\mu$, is weighted by the ratio of the two Bessel functions.
\item[\emph{ii)}] Interestingly, model \eqref{eq:y}-\eqref{eq:th} can be straightforwardly extended to the compressive framework, by replacing the Gaussian prior model by a Bernoulli-Gaussian one. Doing so, a compressive version of the procedure can be simply obtained by adding an update equation of the distribution on the sparse-representation support as proposed in \cite{Dremeau2012}.
\end{itemize}

Coming back to problem \eqref{eq:pb}, an approximation of $p(\xv|\yv)$ then simply follows from
\begin{align}
p(\xv|\yv) \simeq q(\xv) = \prod_i q(x_i),
\end{align}
resulting in the estimation $\hat{\xv}$ such that, $\forall i\in\lbrace1,\ldots,N\rbrace$,
\begin{align}
\hat{x}_i = \argmax_{x_i} q(x_i) = m_i.
\end{align}
In this way, the MF approximation leads finally to a solution which
tends to maximize the posterior probability of \emph{each} $x_i$
(rather than of the whole vector $\xv$). In fact (see
\cite{Dremeau2012,6875083} for a similar discussion) at each
iterations, the KL divergence decays, and since it is bounded, the
algorithm must converge.

\subsection{Noise estimation}
\label{subsec:noise}
As emphasized in \cite{Dremeau2012}, the estimation of model parameters, and in particular the noise variance, can easily be embedded within the VB-EM procedure \eqref{eq:qt}-\eqref{eq:resi2}. Particularized to model \eqref{eq:y}-\eqref{eq:th}, this leads to
\begin{align}
\hat{\sigma}_n^2\negmedspace&= \argmax_{\sigma^2_n}\negmedspace \int_{\tv}\negmedspace \int_{\xv}q(\tv)q(\xv)\log p(\xv,\tv,\yv\vert \sigma^2_n)  d\xv\, d\tv ,\label{eq:estimvar}\\
	         &=\frac{1}{M}\; \left(\;\yv^H\yv-2\;\Re(\bar{\yv}^H\sum_i m_i\dv_i) +\sum_i \Sigma_i\;\dv_i^H\dv_i\right. \nonumber\\
	         &\left.\quad\quad\quad\quad\quad\quad\quad\quad\quad+\sum_i\sum_{j} m_i^*\;m_j\;\dv_i^H\dv_j\right). \label{eq:estimvar2}
\end{align}

Although it may not seem necessary if the noise is known, the
iterative estimation of the noise variance is in fact fundamental and
improves the behavior of the algorithm that would be, otherwise,
trapped close to poor solutions. This point, that was discussed in
detail in \cite{6875083} in the case of compressed sensing, can be
intuitively understood by noticing that $\hat{\sigma}_n^2$ is a
measure of the (mean) discrepancies between the observations and the
assumed model.

\section{Experiments}
\label{sec:experiments}

In this section, we study the performance of the proposed algorithm by
extensive computer simulations. We evaluate and compare the
performance of $4$ different algorithms: Gerchberg-Saxton
\cite{Gerchberg1972}, PhaseCut \cite{Waldspurger2013} and the
variational Bayesian approaches: prGAMP introduced in
\cite{Schniter2012} and considered here in its non-compressive
version, and the proposed MF-based one, denoted by prVBEM. The
algorithms present different complexities. The implementation of
PhaseCut relies on interior-point methods, with a complexity growing
as $\Os(M^{3.5}\log(1/\epsilon))$ where $\epsilon$ is the target
precision \cite{Waldspurger2013}. Gerchberg-Saxton, prGAMP and prVBEM
share similar complexities, of order $\Os(M^2)$ \emph{per iteration}.
We use for PhaseCut and prGAMP the implementations proposed by their authors on their webpages\footnote{resp. http://www.di.ens.fr/data/software/ (released June 18th, 2012) and http://gampmatlab.wikia.com/wiki/Generalized\_Approximate\_Message\_Passing (released May 21st, 2014)}. Within these implementations, the stopping criterium is set to its default value for prGAMP; after preliminary testing and considering the best tradeoff between recovery performance and computational time, Gerchberg-Saxton stops after $3000$ iterations and PhaseCut is run until the target precision drops below $10^{-2}$. Finally prVBEM iterates as long as the KL divergence \eqref{eq:KL} decreases more than a difference of $10^{-8}$ between two consecutive iterations.

In the sequel, we emphasize two desirable properties of the proposed
algorithm. To that end, we consider the following general experiment
setup. Observations are generated according to model
\eqref{eq:y}-\eqref{eq:th} with the following parameters: $N=64$,
$M=\alpha N$ with $\alpha\in\lbrace 1,\ldots, 6\rbrace$ and
$\sigma^2_x=1$. The elements of the dictionary $\D$ are \emph{i.i.d.}
realizations of a zero-mean circular Gaussian distribution with
variance $M^{-1}$. For a fair comparison of the algorithms, prGAMP and
prVBEM are both not aware of the values of $\sigma^2_n$ and
$\sigma^2_x$ used to generate the observations.

We assess the performance in terms of the reconstruction of the signal
$\xv$. In particular, we consider the correlation between the
estimated signal and the one used to generate the data, as a function
of the number of measurements $M$. This figure of merit is evaluated
from $100$ trials for each simulation points.

\subsection{Robustness to noise}
\label{subsec:robustness}
\begin{figure*}[t]
\begin{center}
	\subfigure[]{\includegraphics[width=0.65\columnwidth]{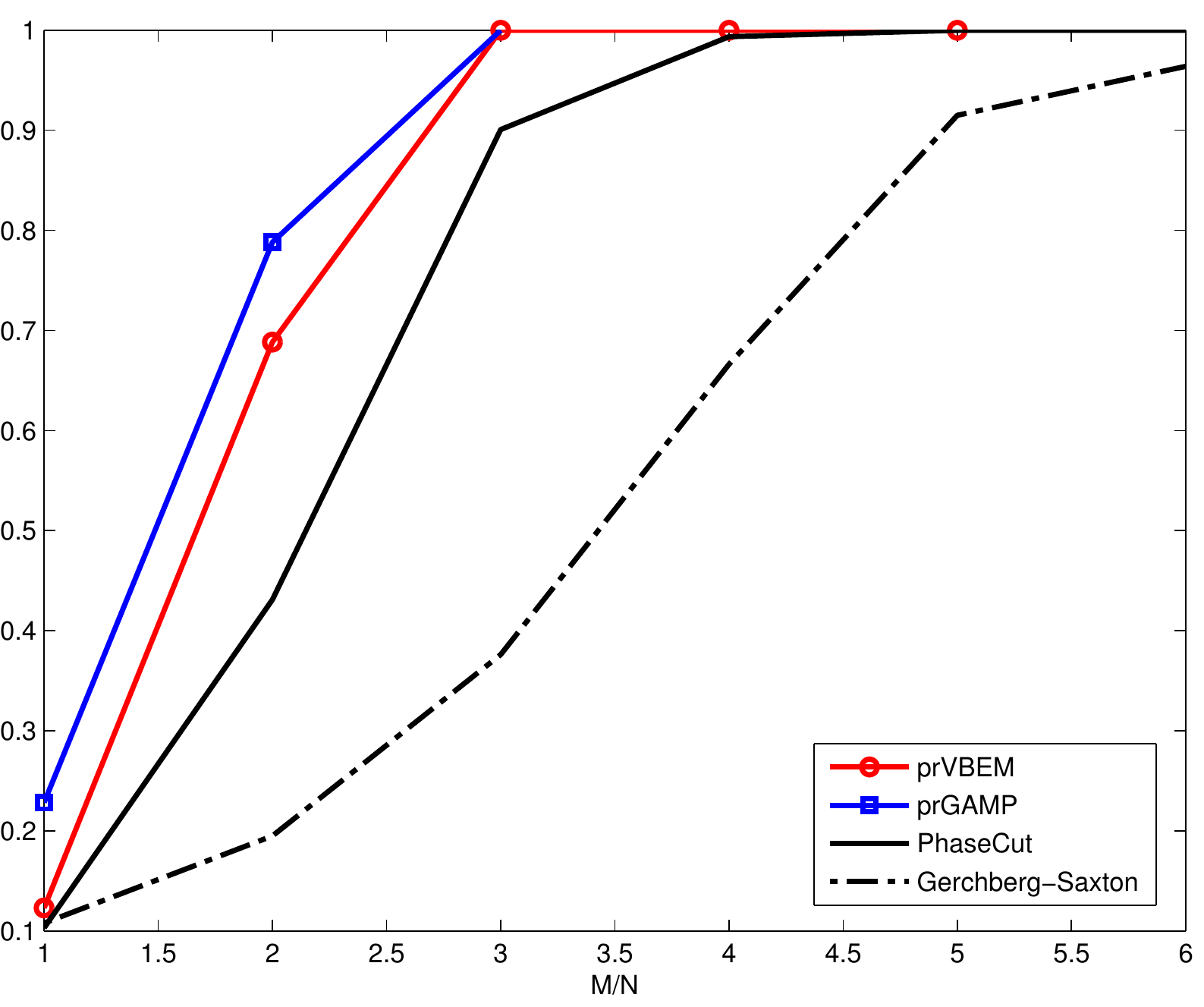}}\hspace{0.1cm}
	\subfigure[]{\includegraphics[width=0.65\columnwidth]{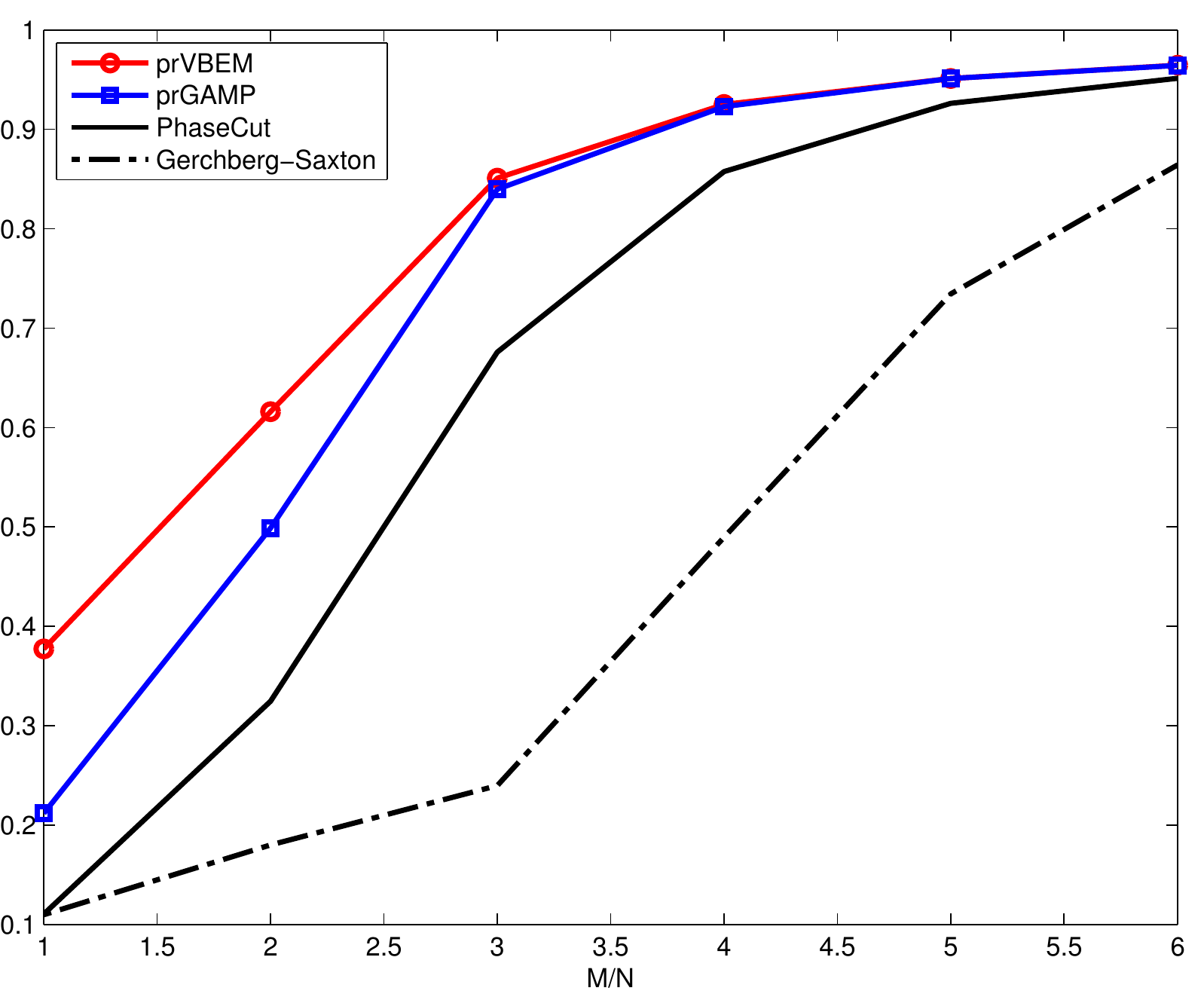}}\hspace{0.1cm}
	\subfigure[]{\includegraphics[width=0.65\columnwidth]{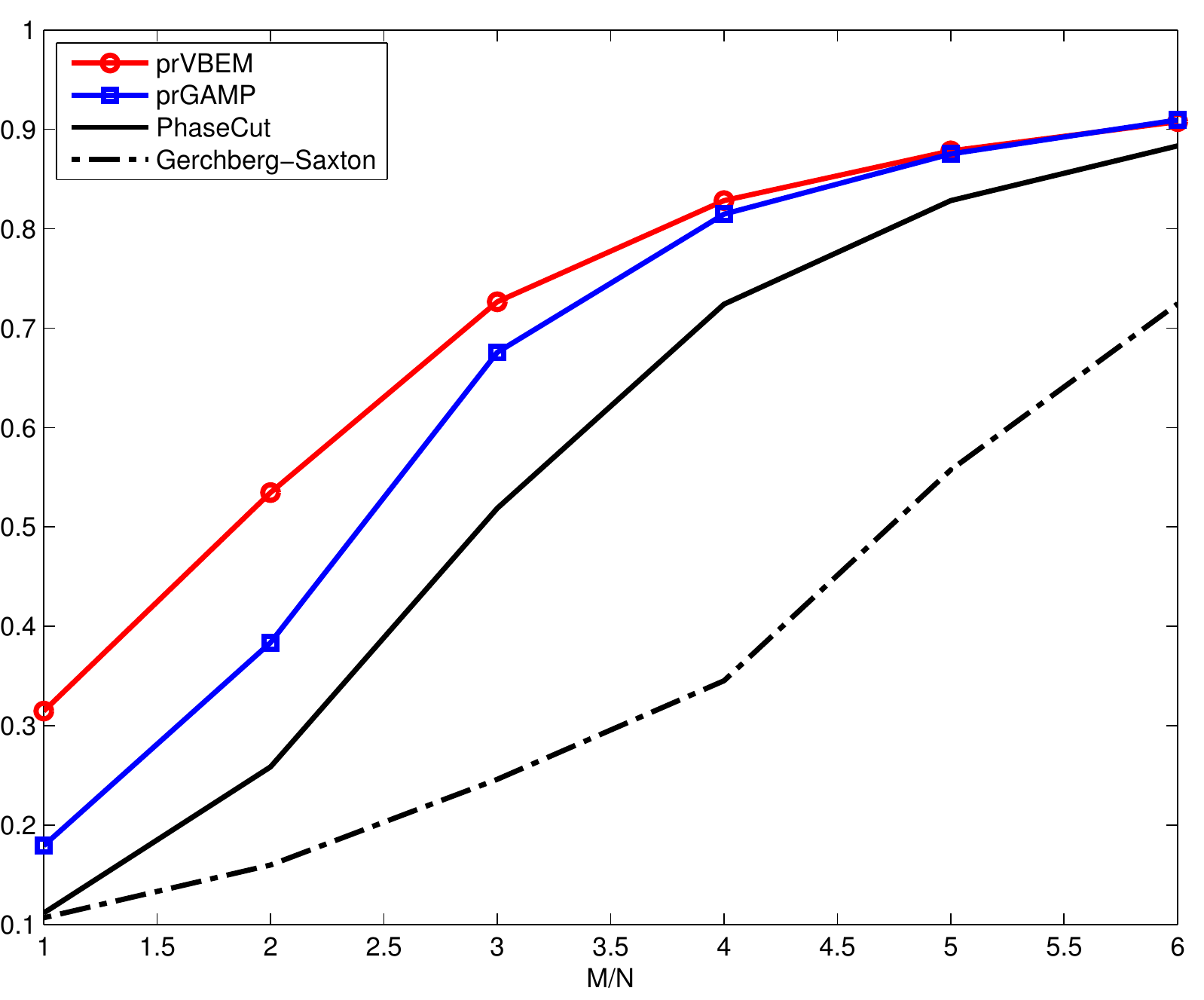}}
\caption{Correlation between estimated and original signals, as a function of the number of measurements $M$ (x-axis is $M/N$ with $N=64$): noiseless case (a), $\sigma_n^2=0.3$ (b) and $\sigma_n^2=0.7$ (c).}\label{fig:corr}
\end{center}
\end{figure*}

We first compare the performance of the algorithms in three different contexts: the noiseless case, \ie $\sigma_n^2=0$ (Fig. \ref{fig:corr}(a)), and two noisy cases where $\sigma_n^2=0.3$ (Fig. \ref{fig:corr}(b)) and $\sigma_n^2=0.7$ (Fig. \ref{fig:corr}(c)). 

As we can see in Fig. \ref{fig:corr}(a), the noiseless case is an ideal case: within the considered setup, as soon as $M>3N$, almost all algorithms obtain a correlation higher than $0.9$. Exception is made by the Gerchberg-Saxton algorithm, which requires in this setup at least $5N$ measurements to perform an acceptable reconstruction.

Most of practical experimental schemes however lead to very noisy measurements. In such scenarios, the phase recovery may become difficult. This is illustrated in Fig. \ref{fig:corr}(b) and (c) where we can observe a general degradation of the performance. Beyond this rough tendency, we note interestingly that a gap seems to open up between the performance obtained by prVBEM and the other algorithms. Although the proposed algorithm does not reach a correlation equal to $1$, it appears relatively more robust and outperforms all other algorithms.

\subsection{Computational time}
\label{subsec:time}
\begin{figure}[h!]
\begin{center}
	\includegraphics[width=0.8\columnwidth]{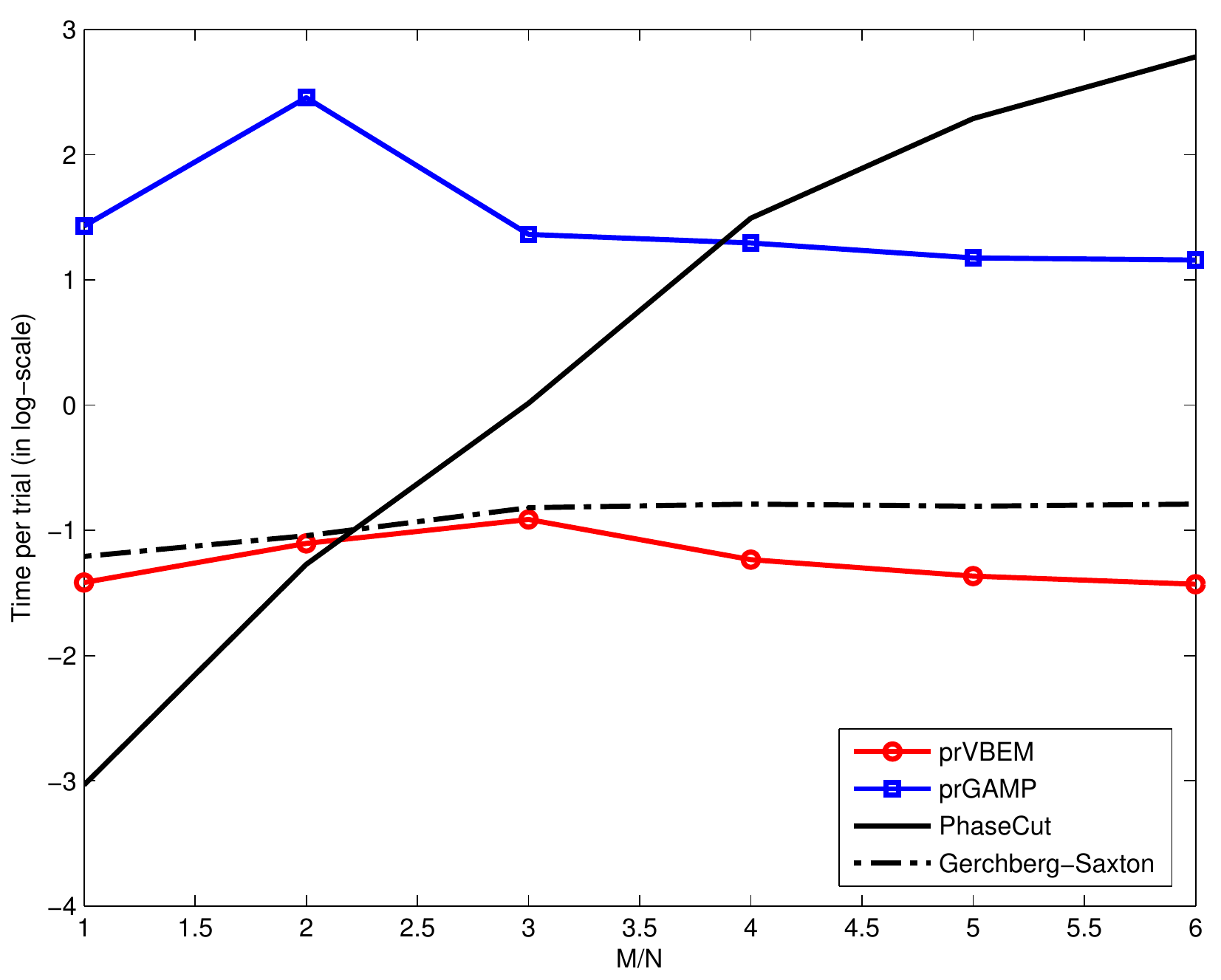}
\caption{Average running time (log-scale) as a function of the number of measurements $M$ (x-axis is $M/N$ with $N=64$).}\label{fig:time}
\end{center}
\end{figure}
Considering the previous noiseless setup, we focus then on the computational costs of the different algorithms. In Fig. \ref{fig:time}, we see that prVBEM presents the lowest running time, in particular lower than prGAMP while sharing a similar complexity order per update step.
It is of course difficult to compare the running times of algorithms which do not have the same stopping criteria. However, this comparison highlights the general good convergence properties of the VB-EM algorithm, which, in that phase-recovery case, requires less than $300$ iterations to significantly decrease the KL divergence \eqref{eq:KL}. When dealing with high-dimensional data, this behavior may be highly desirable.

We refer the interested reader to the author's
webpage\footnote{http://angelique.dremeau.free.fr} for the
implementation of prVBEM.

\section{Conclusion}
\begin{sloppypar}
Using the classical variational mean-field approach to Bayesian
inference, we have presented a new iterative algorithm to
estimate complex vectors from the moduli of their linear
projections (the so-called phase retrieval problem). This leads to a
principled, convergent, and efficient procedure. As far as our simulations are concerned, we have shown in particular that the algorithm performs very well,
both in terms of running time and stability with respect to the state-of-art. A natural
continuation would be to consider the sparse compressive case, and to
apply our approach to relevant experimental situations, such as, for
instance, optical ones \cite{liutkus2014imaging}.
\end{sloppypar}

\section{Acknowledgements}
\label{sec:merci}
This work has been supported in part by the ERC under the European
Union's 7th Framework Programme Grant Agreement 307087-SPARCS. 
AD is currently working at ENSTA Bretagne, Lab-STICC (UMR 6285), 2 rue Francois Verny, F-29200 Brest, France.


\newpage

\bibliographystyle{IEEEbib}
\bibliography{Biblio}

\end{document}